%% file: samplepaper.tex
\documentclass[runningheads]{llncs}
\usepackage{graphicx}
\usepackage{color}
\usepackage{dirtytalk}
\usepackage{csquotes}
\usepackage{multirow}
\usepackage[table,xcdraw]{xcolor}
\begin{document}
\title{Natural interaction with traffic control cameras through multimodal interfaces}
\titlerunning{Natural interaction w. traffic control cameras through multimodal interfaces}
%
\author{Marco Grazioso\inst{1}\orcidID{0000-0002-4056-544X} \and
Alessandro Sebastian Podda\inst{2}\orcidID{0000-0002-7862-8362} \and
Silvio Barra\inst{1}\orcidID{0000-0003-4042-3000} \and
Francesco Cutugno\inst{1}\orcidID{0000-0001-9457-6243}}
\authorrunning{M. Grazioso et al.}
%
\institute
{Department of Electric and Information Technology Engineering (DIETI), University of Naples ``Federico II" \email {\{marco.grazioso,silvio.barra,francesco.cutugno\}@unina.it}\and
Department of Mathematics and Computer Sciences, University of Cagliari
\email{sebastianpodda@unica.it}\\}
\maketitle       
\begin{abstract}
Human-Computer Interfaces have always played a fundamental role in usability and commands' interpretability of the modern software systems. With the explosion of the Artificial Intelligence concept, such interfaces have begun to fill the gap between the user and the system itself, further evolving in Adaptive User Interfaces (AUI). Meta Interfaces are a further step towards the user, and they aim at supporting the human activities in an ambient interactive space; in such a way, the user can control the surrounding space and interact with it. This work aims at proposing a meta user interface that exploits the \emph{Put That There} paradigm to enable the user to fast interaction by employing natural language and gestures. The application scenario is a video surveillance control room, in which the speed of actions and reactions is fundamental for urban safety and driver and pedestrian security. The interaction is oriented towards three environments: the first is the control room itself, in which the operator can organize the views of the monitors related to the cameras on site by vocal commands and gestures, as well as conveying the audio on the headset or in the speakers of the room. The second one is related to the control of the video, in order to go back and forth to a particular scene showing specific events, or zoom in/out a particular camera; the third allows the operator to send rescue vehicle in a particular street, in case of need. The gestures data are acquired through a Microsoft Kinect 2 which captures pointing and gestures allowing the user to interact multimodally thus increasing the naturalness of the interaction; the related module maps the movement information to a particular instruction, also supported by vocal commands which enable its execution. Vocal commands are mapped by means of the \texttt{LUIS} (Language Understanding) framework by Microsoft, which helps to yield a fast deploy of the application; furthermore, \texttt{LUIS} guarantees the possibility to extend the dominion related command list so as to constantly improve and update the model.
A testbed procedure investigates both the system usability and multimodal recognition performances. Multimodal sentence error rate (intended as the number of incorrectly recognized utterances even for a single item) is around 15\%, given by the combination of possible failures both in the ASR and gesture recognition model. However, intent classification performances present, on average across different users, accuracy ranging around 89-92\% thus indicating that most of the errors in multimodal sentences lie on the slot filling task. Usability has been evaluated through task completion paradigm (including interaction duration and activity on affordances counts per task), learning curve measurements, \textit{a posteriori} questionnaires.


\end{abstract}
\input{sections/01_introduction}

\input{sections/02_Related_Works}

\input{sections/03_System_Architecture}

\input{sections/04_System_Evaluation}

\input{sections/05_Conclusions}

\bibliography{refs.bib}
\bibliographystyle{splncs04}
\end{document}

%% file: sections/01_introduction.tex
\section{Introduction}
The user experience and the user interfaces often make a difference in the choice of a software rather than another. This because, during the years, the usability of a system has become the most powerful evaluation criteria, given the central role of the user within the life cycle of a software \cite{wallach2012user}. As a consequence, the user interfaces have accordingly modified, so to satisfy the user requirements, which suggest more natural interaction means \cite{VUI2020}. Adaptive User Interfaces \cite{browne2016adaptive} are an evolution of the common UI, in which the interface adapts itself to meet the user interaction criteria; such kind of interfaces have furtherly narrowed the gap between the system and the user, since the interaction between the two components improves asymptotically after a partial experience of the system itself with that user \cite{ML4AUI}. A further step forward is identified in the Meta User Interfaces \cite{coutaz2007meta} which improve the UX by adding particular care to the environments which s/he acts within. 
The definition of such kind of interfaces is reported from the work in \cite{coutaz:DSP:2007:1082}:
\begin{displayquote}
\textit{The interface is \textit{meta} because it serves as an umbrella beyond the domain-dependent services that support human activities in this space. It is \textit{UI}-oriented because its role is to allow users to control and evaluate the state of the ambient interactive space. By analogy, a \textit{metaUI} is to ambient computing what desktops and shells are to conventional workstations. }
\end{displayquote}

These interfaces aim at supporting human activities in an ambient interactive space\cite{coutaz2007meta}; in such a way, the user can control the surrounding space and interact with it. Besides being very useful in smart environments \cite{MUI_2009}, given their inner ability to allow interaction with the surrounding space, Meta User Interfaces develop their usefulness in contexts in which speed of action and reaction are fundamental, like in surgery scenarios \cite{rosa2014use,sanchez2017use} or in scenarios in which the user needs to have the entire surrounding under direct control.

In this paper, a Meta User interface for a video surveillance control room application scenario \cite{Atzori2021} is proposed; the interface is based on the \emph{Put That There} paradigm \cite{bolt1980put}, so to enable the user to a fast interaction by using natural language and gestures. Three interaction environments are considered:
\begin{itemize}
    \item the \textit{control room environment}: in this environment, the user is given both the ability to organize the views of the monitor s/he controlling, and the option to convey the audio of a specific monitor towards the headset or in the speakers spread into the room;
    \item the \textit{video management environment}: the user can navigate a particular video, so to reach a specific minute or a particular scene showing a specific event. Also, s/he can zoom in/out a particular screen as well as applying a split-screen to compare a specific trait of the road from to different points of view (if proper ground cameras are placed). Finally, also the possibility to pan, tilt and zoom a particular camera is provided (if the camera is provided with such mechanics);
    \item the \textit{road}; the user is offered the skill to act immediately whether an intervention is required on a road; in such sense, the operator is provided with interaction means for sending rescue vehicle in a particular street, in case of need.
\end{itemize}

The main contributions of the proposed paper are the following:
\begin{itemize}
    \item an entity-extensible tool for gestures and vocal interaction with the surrounding environment;
    \item three environments act as the object of the interaction: the control room, the displayed video and the surveilled road;
    \item the system exploits the Kinect for modelling the operator joints and the \texttt{FANTASIA} framework for the rapid development of interactive applications.
\end{itemize}
The remainder of the work is organized as follows: Section \ref{rw} discusses the state of the art; Section \ref{sa} describes the entire system architecture and the related modules for gesture interactions and voice commands. Section \ref{ev} evaluates the system and reports details about the testbed. Finally, Section \ref{conc} concludes the paper and explores future directions.

%% file: sections/02_Related_Works.tex
\section{Related work} \label{rw}
The user interfaces are nowadays strongly oriented to the improvement of the user experience, especially in those factors related to accessibility, adaptability and control. The accessibility finds its leading exponent in the multimodal interfaces \cite{oviatt2003multimodal}, which provide several modalities of interaction with the system, thus resulting useful not only for normal users, which are able to choose the preferred interaction mode \cite{8564178}, but also for people with physical impairments whose interaction criteria are met by one or more of the provided modalities \cite{10.1007/978-3-319-93846-2_73}. Few examples are described in \cite{Rocha:2020aa} and in \cite{paudyal2020voiceye}: both are oriented to disabled users, in order to facilitate their interaction without using mice or keyboards; the first proposes an augmented communication system whose interface is controlled by different types of signals, like electromyography, electrooculography and accelerometer. In the second, instead, Voiceye is described, a multimodal interface system that combines voice input and eye-gaze interaction for writing code. Multimodal interfaces are also very useful in those environments in which the user needs his/her hands for primary tasks, and therefore the interaction must take place in other ways. As an example, drivers have their hands on the steering wheel and therefore interactions with the surrounding cockpit must happen in other ways. An interesting multimodal interaction system oriented to drivers is presented in \cite{9118888} which along with the common interaction modalities like touch and physical button, the authors proposes further modalities like \textit{mid-air gestures}, \textit{voice} and \textit{gaze}.
\\
AI-driven interfaces are more reliable for those systems which need to modify themselves during the interaction in order to further adapt to the user him/herself. An example of systems is described in \cite{10.1145/3379336.3381467} in which the application proposed delivers personalized and adaptive multimedia content tailored to the user. Other examples of adaptive systems are located in the field of robotics and Human-Robot Interaction, like in \cite{wijayasinghe2018adaptive} and in \cite{dosadaptive}. In \cite{10.1007/978-3-030-51328-3_4} an AI-driven approach for additive manufacturing is proposed. Many systems are also aimed at improving learning platforms, like in \cite{kim2020ai} in which the engagement and motivation of the students are inferred by analysing their implicit feedbacks.
\\
A further step towards the complete environment automation and complete control for the user is given by the meta user interfaces whose main focus is to support the interaction with ambient intelligence systems \cite{MUI2017}. Such kind of interaction earns much importance in the current era of IoT and smart environments, like highlighted in \cite{IoTInteraction}, in which the authors describe a 3D-based user interface for accessing smart environments. In particular, many works have dealt with the problem of interacting with Smart Homes \cite{BALTAOZKAN2013363,whatwouldyouoask}, since, in such contexts, the user needs to control not a single item, but in some cases, the objects of the requirements are a bunch of sensors which have to cooperate in order to produce a result. From this point of view, there have been many works dealing with such issues \cite{designandevaluation} and the answers have been very controversial; in fact, in order to assist the user in the best way possible, some systems need to design both machine-to-human and machine-to-machine (M2M) communication systems \cite{7164704}.

%% file: sections/03_System_Architecture.tex
\section{System architecture} \label{sa}
\begin{figure}[!ht]
\includegraphics[width=\textwidth]{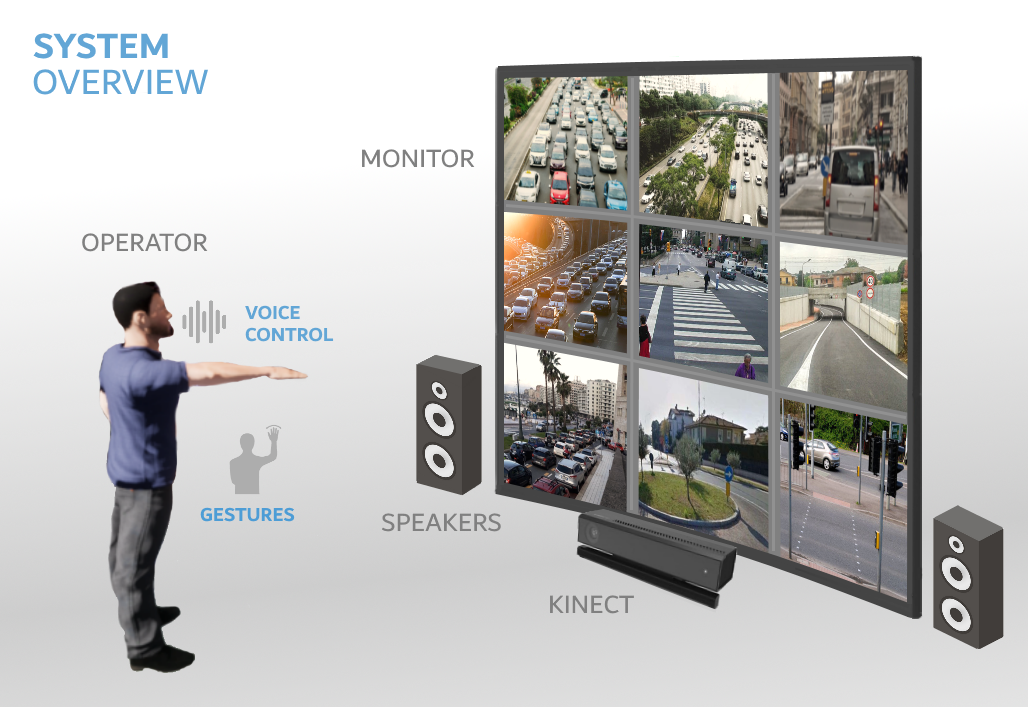}
\centering
\caption{The figure illustrates the application scenario. Note that the operator has a widescreen at his disposal, which he can interact with; the screen is logically divided in a 3-by-3 matrix-shaped form. At the basis of the monitor, the Kinect gathers the gestures of the operator. The speakers can be placed within the room, independently by the position of the monitor. Also, an environmental microphone (or a headset) gets the vocal commands of the operator.}
\label{fig:scenario}
\end{figure}

The applications scenario is shown in Figure \ref{fig:scenario}; it depicts the control room in which the operator is immersed.
The operator is located about 2 meters from the monitor, so as to have a clear overall view of all the supervised streets. The interactive area consists of a 2,5 meters high and 4,4 meters long curved screen, which nine street camera views are displayed on in a $3-by-3$ configuration.

On the floor, on the basis of the screen, a Microsoft Kinect 2\cite{zhang2012microsoft} is placed, which captures user pointing and gestures so as to allow him a very natural interaction; simultaneously, a microphone capture the vocal command uttered by the user. The scenario is completed by a set of environmental speakers located in the upper corners of the room; also, the user can wear a headset provided with the microphone.

The gestures data are acquired through the related module that maps the movement information to a particular instruction, also supported by vocal commands which enable its execution. The videos are synthetically generated and acquired from the \textit{Grand Theft Auto V} video-game by Rockstar Games, by using a particular patch that enables specific tricks within the video-game, like the custom positioning of the cameras within the scene, creation of pedestrians for simulating crowds, traffic lights control and so on. Vocal commands are mapped by means of the \texttt{LUIS} (Language Understanding) framework by Microsoft, which helps to yield a fast deploy of the application; moreover, \texttt{LUIS} guarantees the possibility to extend the domain-related command list, to constantly improve and update the model. This framework receives a previously defined set of commands as input; there is no linguistic constraint to the possible utterances the user can produce, thus leading to a very natural way of producing commands.
Speech and video features are acquired asynchronously and thus, a step of data fusion is necessary to perform the interpretation of the commands and make the system able to accomplish the users' requests. 
It has been adopted a multimodal fusion technique that makes it possible to easily incorporate expert knowledge and domain-specific constraints within a robust, probabilistic framework: OpenDial \cite{lison2016opendial,Grazioso2020}. Here, the input signals are modeled into random variables and the system state is represented as a Bayesian Network, where probabilistic rules define the way the system state is updated. Entities recognised into the speech channel compose the probabilistic rule together with labels signalling the gesture instances.

The interactive 3D environment projected on the curved screen has been developed by using the popular game engine \emph{Unreal Engine 4}\footnote{https://www.unrealengine.com/} and by taking advantage of the facilities provided by the framework \texttt{FANTASIA}\cite{origlia2019fantasia}. The framework aims at supporting a rapid design and implementation of interactive applications for HCI studies by integrating several modules, such as voice synthesis, speech recognition, graph database and dialogue manager.

\subsection{Natural language understanding} \label{nlu}
To gather data to model the different intents, a \emph{Google Form} questionnaire has been released among the contacts of the authors, including operators that daily work in a video surveillance scenario. 25 people have answered the form, with age between 32 and 55 years. The questions aimed at gathering all the utterances which could be used to achieve a specific operation in the depicted context. 
An example of question has been related to all the possible ways to refer to a specific monitor. Recall that the screen is divided into nine monitors in a 3-by-3 shaped matrix. In this context, the single monitor can be referred in different ways; following some of the answers obtained:
\begin{itemize}
    \item upper-right monitor - monitor in (1,1) - first monitor - north/west monitor;
    \item central monitor - monitor in (2,2) - monitor at the center - fifth monitor;
    \item lower-left monitor - monitor in (3,3) - last monitor - monitor number 9;
    \item ...
\end{itemize}

Analysing the utterance, a recurrent pattern emerges. In most of cases, people start the sentence by declaring the operation to do, followed by the target (or targets) of such operation. This observation has guided our choice for applying the intent-entity paradigm. This paradigm involves two steps: intent classification and entity recognition. The first task is in charge of the identification of the general purpose of the user query, e.g. the operation to be executed. The second one is responsible for retrieving from the query the objects which have a meaning in the specific domain. Typically, these tasks require the developing of neural networks and, as a consequence, the need to access a huge amount of labelled data; hence, starting from scratch is not a feasible solution. Therefore, the idea has involved the use of \texttt{LUIS}, so to take advantage of its capabilities. \texttt{LUIS} is a module included in the Microsoft cognitive services, which provides a simple way of loading sentences, defining intents and entities for data annotation, training models, even with a small amount of data, and exposing it as a service.
Moreover, it gives the chance to define complex entities composed of a set of smaller sub-entities. This level of representation allows us to define more general concepts that could be composed differently, based on the contained sub-entities. For instance, the general concept \textit{monitor} could be composed by the couple $<$\textsc{monitor, reference number}$>$ or by the pair \textsc{(x,y)} like in a matrix.

Starting by the matrix view shown in Figure \ref{fig:scenario}, the following intents have been identified:
\begin{itemize}
    \item \textsc{zoom\_in - zoom\_out}: for enlarging a specific monitor (zoom in) or for going back to the matrix view (zoom out);
    \item \textsc{split\_screen}: the matrix view is substituted by a view in which two videos are placed side by side;
    \item \textsc{swap}: swap the position on two videos into the matrix view; 
    \item \textsc{audio\_to\_device}: the current audio is conveyed to a certain device (a headset or the room speakers);
    \item \textsc{audio\_off}: switch off the audio;
    \item \textsc{rewind - forward}: go back or forth to a particular minute in the video.
\end{itemize}

The involved entities are the monitor, which is composed by the sub-entities \textit{ref}, \textit{ref\_x} and \textit{ref\_y}, and the device, in order to capture the reference to an audio device; in addition also deictic terms are modelled, so to allow the user to utter expressions like \textit{this} and \textit{that} (as an example, for referring to a particular monitor), in their singular and plural form.

    \subsection{Pointing recognition} \label{pr}
\begin{figure}[!ht]
        \includegraphics[width=0.5\textwidth]{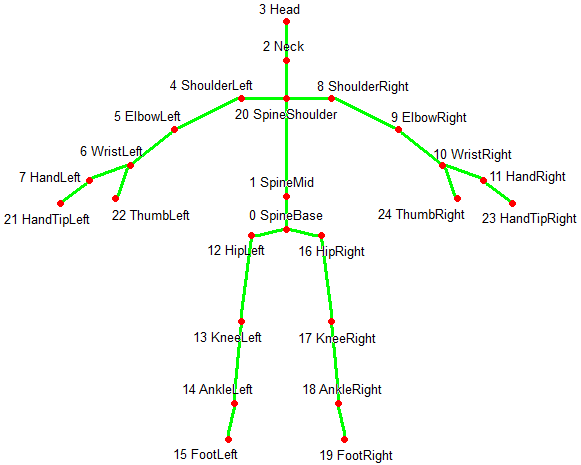}
        \centering
        \caption{Joints provided by the Kinect\cite{Ousmer2019}}
        \label{fig:kinect_joints}
    \end{figure}

    The pointing recognition module acts in an independent way, by asynchronously collecting the skeleton information provided by the Kinect sensor. Skeleton data consists of a set of 3D points representing the user's joints (see Figure \ref{fig:kinect_joints}). The coordinates of such points refer to the Kinect coordinates system, where the axes origin corresponds to the sensor position. Since the 3D environment coordinates system does not match the Kinect one, the skeleton data has been transformed by rotating them according to the sensor inclination, so as to properly representing it in the 3D environment. Moreover, the user height, the distance from the Kinect and the lateral displacement are taken into account.
     Skeleton data, in combination with the Skeletal Mesh object provided by Unreal Engine, could be used to spawn an avatar of the user in the virtual space. The Skeletal Mesh consists of a hierarchical set of interconnected bones and it gives the chance to associate its joints with the Kinect one, obtaining a virtual representation of the user that follows his movements.
     Once obtained a good user representation, the next step is to estimate where the user is pointing at. This process could be divided into two sequential tasks:
     \begin{itemize}
         \item Pointing detection.
         \item Pointing recognition.
     \end{itemize}
     In pointing detection, it is important to distinguish between pointing and non-pointing gestures. Since, at the moment, the system does not recognize other kind of gestures, it is possible to use the hand position and movement speed as discriminant. In particular, by computing the distance between the SpineMid joint and the hand one, such positions where the hand is very high or very low could be excluded, assuming that the user is not pointing to the screen. Moreover, an high speed movement of the hand suggests that the system is observing a transition movement and it must be excluded, too. Exclusion criteria are based on fixed thresholds empirically estimated.
     The detected gestures can now be processed to recognize the pointed object. To accomplish this task, a geometrical approach is used: it computes the line passing through the shoulder joint and the hand one, and stretches it forward until it collides with an environment object. In order to avoid errors caused by possible noise in the joints data and, eventually, transition movements that passed the first filtering step, our approach collects the pointed objects inside a window of 1 second and then, for each different object, computes the probability to be the current pointed object.
    
    \subsection{Multimodal Fusion Engine} \label{mf}
    The Natural Language Understanding (NLU) and the pointing Recognition activities, discussed in the previous subsections, have been fused into the Multimodal Fusion Engine here discussed, in order to provide the proposed concept of multimodality. Following the suggestions discussed in \cite{Oviatt2000}, the Multimodal Fusion Engine has been developed as an independent module. It receives asynchronous messages from the input modules related to the NLU and the gesture recognition, handled by the specific receivers:
    \begin{itemize}
    \item the NLU message consists of the sentence uttered by the user, together with the related intents and entities. Also, a confidence value is returned related to the recognition of the intents and the entities. 
    \item the gesture recognition message consists of the pointed ojects, together with the related confidence values;
    \end{itemize}
    
    The OpenDial framework is in charge of receiving and managing the inputs from both the gesture recognition and the NLU modules. OpenDial has the peculiarity to be capable of managing the dialogue using a mixed approach based on both rule-based and probabilistic approaches. Indeed, whether it allows to integrate expert knowledge by defining rules, on the other hand it allows probabilistic reasoning by defining Bayesian networks. 
    Given that, the received messages are mapped to the respective random variables encoded in a Bayesian network, so to derive a common interpretation.
    During the multimodal fusion, several aspects need to be considered and modelled to avoid wrong interpretation. According to the paper in \cite{Grazioso2020}, several OpenDial modules have been developed, that change the network configuration according to specific constraints.
    The \emph{Multimodal Input Integrator} module aims at combining input variables coherently. In particular this module analyses verbal actions and pointed objects in order to understand the current request. Since the variables evolve in real-time, the \emph{Multimodal Time Manager} is used to check the consistency and prune out-of-date variables. In particular, starting from time-stamps related to the input variables, once a new speech signal is captured, the module compares its time intervals with those computed for each pointing variable, pruning off pointing gestures whose occurrence was concluded more than 4 seconds before the start of the current speech signal. Pruning criteria were selected in accordance with the study discussed in \cite{SynchMultimodal97}.
    In order to support multi target operations, (as an example "\textsc{swap this monitor with this one}) the system needs to keep in memory more than the last pointed object; to this regard, a linked list is implemented, so to keep trace the pointed objects from the most recent one to the last one respecting the previous criteria.
    Since the input variables come asynchronously, the \emph{State Monitor} manages the entire operation by observing changes in dialogue state. Therefore, the unification methods are called by this component according to dialogue progresses. Once the system has derived the current request, the message containing the most appropriate action to be performed is sent to the game engine.

%% file: sections/04_System_Evaluation.tex
\section{Evaluation} \label{ev}
A test bed procedure is used to investigate both the system usability and multimodal recognition performances.
Users have been involved in a task-oriented test consisting of six tasks. Since the main aim is to test the usability of the application in a real scenario, each of the tasks is composed of both a voice command and a gesture activity. In such a way all the involved modules are tested in their entirety.

After a brief introduction on both the scenario and the interaction modality, the task list was given to the users and they were then left free to interact as they preferred to accomplish their jobs. During the tests, users have been recorded in order to analyze their behaviour, so to obtain usability information. Each task was annotated with one of following labels: \textsc{s} for success, \textsc{ps} for partial success and \textsc{f} for failure. The explanation of each outcome is explained as follows:
\begin{itemize}
    \item Success (\textsc{s}): the user has completed the task acting in an autonomous way and in reasonable times.
    \item Partial Success \textsc{ps}: the user has needed some suggestions or spent more time to complete the task;
    \item Failure \textsc{f}: the user was completely unable to proceed in the task completion.
\end{itemize}
Moreover, the fusion engine logged the information about the fusion process in order to compare its result with the recorded video, so obtaining an estimation of the precision and reliability of the fusion basing on the NLU and Gesture recognition inputs.
\paragraph{The tasks}
How cited above, the tasks have been built such that both the NLU module and the Gesture Recognition module are activated, so to properly execute the Multimodal Fusion Engine.
Before starting a task, nine videos at the same time are shown, organized as a 3-by-3 matrix. The users are placed at 3m from the screen. The Kinect is placed at the bottom of the screen. The room is provided with four speakers at the top corners; however, the user is provided with a wireless headset with a microphone. In Table \ref{tab:tasks} the 6 tasks are reported; however, these have been defined to the user by not using keywords, like \textsc{zoom}, \textsc{swap} and so on, but paraphrases are used. 
\begin{table}[!ht]
\caption{The tasks involved in the test; the users are asked to accomplish the defined actions in sequence. When the action involves a ''Random Monitor'' this means that the user is free to choose the preferred monitors as objects of his/her action.}
\begin{tabular}{|l|l|l|l|l|}
\hline
            & \textbf{Action 1}                                       & \textbf{Action 2} & \textbf{Action 3}                                       & \textbf{Action 4} \\ \hline
\textbf{T1} & \textit{Zoom \textless{}Monitor 1\textgreater{}}        & \textit{Zoom out} & \textit{Zoom \textless{}Monitor 9\textgreater{}}        & \textit{Zoom out} \\ \hline
\textbf{T2} & \textit{Zoom \textless{}Rand. Monitor\textgreater{}}    & \textit{Zoom out} & \textit{Zoom \textless{}Rand. Monitor\textgreater{}}    & \textit{Zoom out} \\ \hline
\textbf{T3} & \textit{Split \textless{}Monitors (1, 9)\textgreater{}} & \textit{Zoom out} & \textit{Split \textless{}Monitors (3, 7)\textgreater{}} & \textit{Zoom out} \\ \hline
\textbf{T4} & \textit{Split \textless{}Rand. Monitors\textgreater{}}  & \textit{Zoom out} & \textit{Split \textless{}Rand. Monitors\textgreater{}}  & \textit{Zoom out} \\ \hline
\textbf{T5} & \textit{Swap \textless{}Monitors (1, 9)\textgreater{}}  & \textit{Zoom out} & \textit{Swap \textless{}Monitors (3, 9)\textgreater{}}  & \textit{Zoom out} \\ \hline
\textbf{T6} & \textit{Swap \textless{}Rand. Monitors\textgreater{}}   & \textit{-}        & \textit{Swap \textless{}Rand. Monitors\textgreater{}}   & \textit{-}        \\ \hline
\end{tabular}
\label{tab:tasks}
\end{table}

Twelve participants have been hired for the testing phase; three considerations need to be done, in order to highlight the fairness of the test.:
\begin{itemize}
    \item none of the participants works as operator in video surveillance field;
    \item all of the participants can be defined as average computer users;
    \item none of the participants have been invited to fill the form used for defining entities and intents.
\end{itemize}

\subsection{Results} \label{res}
The data collected during the interaction between the system and the users have been used to generate the results in Table \ref{table:task_completion}. This analysis represents a simple way of estimating the system usability by computing the \textit{task completion rate}. This measure has been computed by counting the total successes, the partial successes and the failures of the users in completing the assigned task and by making a weighted average of those values. In particular, the success has a weight equal to $1$, the partial success has a weight equal to $0.5$ and failures has a weight equal to $0$. Proceeding as described, a total of 52 successes, 16 partial successes and 4 failures were obtained. Computing the task completion rate, a value of $0.83$ emerged. Considering the few data used to train the NLU model this represents a good result; indeed, analysing the recorded test sessions, it was noticed that the most frequent cause of failure comes from the wrong intent or entity interpretation. This observation suggests that by increasing the amount and the variety of the examples used to train the model, it is possible to improve the results.
Moreover, the success rate for both the NLU and Gesture recognition modules is computed. For the NLU model, the number of correct interpretation over the total speech interaction was counted, providing a success rate of 0.76. As said before, this value is strongly influenced by the amount and the variety of the examples. It is also important to say that answering an online survey is different from really interact with a system; in fact, the data could be not representative enough for the case study. In this regard, a pipeline of active learning would increase the NLU success rate and consequently the task completion rate. This activity would help to collect and interpret the misunderstood sentence, so improving the model.
Regarding the Gesture recognition, the number of correct objects recognised by the system over the total interaction has reached an accuracy of $79\%$. As expected, most of the errors occur in the multiple object intent. Since this is a complex activity, several variables may influence the success of the action. In particular, wrong object selection doesn't come from an imprecise computation of the art direction, but comes from the users' movement speed from an object to another. If the movement takes long, the multimodal fusion starts before the user complete the activity. In most cases, this problem has regarded the users not so comfortable with technology. From this observation, it can be deduced that, to satisfy a larger part of the population, it is necessary to improve the recognition of multi-target intents by providing a time window large enough to consent to end the movement.
It can be concluded that the multimodal fusion works properly under the assumption that both the NLU and Gesture recognition modules do their job correctly.

\begin{table}[!ht]
\caption{Task completion table}
\centering
\begin{tabular}{|c|c|c|c|c|c|c|c|}
\hline
\multicolumn{2}{|c|}{}                                                                                                                               & \multicolumn{6}{c|}{\cellcolor[HTML]{DAE8FC}{\color[HTML]{333333} \textbf{T A S K S}}}                                                                                                                                                                                                                                                                                      \\ \cline{3-8} 
\multicolumn{2}{|c|}{\multirow{-2}{*}{}}                                                                                                             & \cellcolor[HTML]{DAE8FC}{\color[HTML]{333333} \textbf{T1}} & \cellcolor[HTML]{DAE8FC}{\color[HTML]{333333} \textbf{T2}} & \cellcolor[HTML]{DAE8FC}{\color[HTML]{333333} \textbf{T3}} & \cellcolor[HTML]{DAE8FC}{\color[HTML]{333333} \textbf{T4}} & \cellcolor[HTML]{DAE8FC}{\color[HTML]{333333} \textbf{T5}} & \cellcolor[HTML]{DAE8FC}{\color[HTML]{333333} \textbf{T6}} \\ \hline
\cellcolor[HTML]{C0C0C0}                                                                                      & \cellcolor[HTML]{C0C0C0}\textbf{U1}  & S                                                          & S                                                          & S                                                          & S                                                          & S                                                          & S                                                          \\ \cline{2-8} 
\cellcolor[HTML]{C0C0C0}                                                                                      & \cellcolor[HTML]{C0C0C0}\textbf{U2}  & PS                                                         & S                                                          & S                                                          & S                                                          & F                                                          & F                                                          \\ \cline{2-8} 
\cellcolor[HTML]{C0C0C0}                                                                                      & \cellcolor[HTML]{C0C0C0}\textbf{U3}  & S                                                          & S                                                          & S                                                          & S                                                          & S                                                          & S                                                          \\ \cline{2-8} 
\cellcolor[HTML]{C0C0C0}                                                                                      & \cellcolor[HTML]{C0C0C0}\textbf{U4}  & S                                                          & S                                                          & PS                                                         & S                                                          & PS                                                         & PS                                                         \\ \cline{2-8} 
\cellcolor[HTML]{C0C0C0}                                                                                      & \cellcolor[HTML]{C0C0C0}\textbf{U5}  & S                                                          & S                                                          & PS                                                         & S                                                          & PS                                                         & S                                                          \\ \cline{2-8} 
\cellcolor[HTML]{C0C0C0}                                                                                      & \cellcolor[HTML]{C0C0C0}\textbf{U6}  & PS                                                         & PS                                                         & S                                                          & S                                                          & S                                                          & S                                                          \\ \cline{2-8} 
\cellcolor[HTML]{C0C0C0}                                                                                      & \cellcolor[HTML]{C0C0C0}\textbf{U7}  & S                                                          & PS                                                         & S                                                          & S                                                          & S                                                          & S                                                          \\ \cline{2-8} 
\cellcolor[HTML]{C0C0C0}                                                                                      & \cellcolor[HTML]{C0C0C0}\textbf{U8}  & S                                                          & S                                                          & PS                                                         & S                                                          & F                                                          & F                                                          \\ \cline{2-8} 
\cellcolor[HTML]{C0C0C0}                                                                                      & \cellcolor[HTML]{C0C0C0}\textbf{U9}  & S                                                          & S                                                          & S                                                          & S                                                          & PS                                                         & S                                                          \\ \cline{2-8} 
\cellcolor[HTML]{C0C0C0}                                                                                      & \cellcolor[HTML]{C0C0C0}\textbf{U10} & S                                                          & S                                                          & PS                                                         & S                                                          & PS                                                         & S                                                          \\ \cline{2-8} 
\cellcolor[HTML]{C0C0C0}                                                                                      & \cellcolor[HTML]{C0C0C0}\textbf{U11} & S                                                          & S                                                          & S                                                          & S                                                          & PS                                                         & S                                                          \\ \cline{2-8} 
\multirow{-12}{*}{\cellcolor[HTML]{C0C0C0}\textbf{\begin{tabular}[c]{@{}c@{}}U\\ S\\ E\\ R\\ S\end{tabular}}} & \cellcolor[HTML]{C0C0C0}\textbf{U12} & S                                                          & S                                                          & PS                                                         & PS                                                         & S                                                          & S                                                          \\ \hline
\end{tabular}
\label{table:task_completion}
\end{table}

%% file: sections/05_Conclusions.tex
\section{Conclusions and Future work}\label{conc}
In this paper, a meta user interface was proposed. The application scenario of the system is a video surveillance control room, in which an operator has an NLU module and a Gesture recognition module at his disposal, to issue commands to the environments by leveraging both on voice and pointing. The system has involved the use of a Kinect and the \texttt{LUIS} framework for gesture modelling and vocal command processing respectively; the OpenDial framework has been used for fusing information coming from the modules. 
The preliminary results are obtained by assigning six composed tasks to twelve participants; these show that the system is consistently reliable and usable, since the participants were not trained for the test, but they were only explained what the system was intended for. 

\paragraph{Limitations.}
Given the good outcomes of the system results, many advances can be done: first, it would be possible to expand the use of the system also to other purposes, like event tagging, object annotations and so on. This would imply the definition of new entities (\textsc{car}, \textsc{lamppost}, \textsc{pedestrian}, ...) and new intents (\textsc{label}, \textsc{pinch}, ...). This involves an enhancement also in the supporting devices, given the fact that some actions involving the single fingers, and not the hands, cannot be easily recognized by using the Kinect.